\documentclass[5p,sort&compress]{elsarticle}
\usepackage[T1]{fontenc}
\usepackage[utf8]{inputenc}
\usepackage{fourier}
\usepackage{textcomp}
\usepackage{textalpha}
\usepackage{verbatim}
\usepackage{etoolbox}
\usepackage{xinttools}
\usepackage{xspace} 
\usepackage[dvipsnames,svgnames]{xcolor}
\usepackage{amsmath,amsfonts,amssymb}
\usepackage{mathtools}
\usepackage{physics}
\usepackage{siunitx}
\usepackage{miller}
\usepackage{subfig}
\usepackage{subeqnarray}
\usepackage{multirow}
\usepackage{tabularray}
\usepackage{tabularx}
\usepackage{booktabs}
\usepackage[normalem]{ulem}
\usepackage{graphbox}
\usepackage{enumitem} 
\usepackage{calc}
\usepackage{todonotes}
\usepackage{color,soul}
\usepackage{hyperref}

\hypersetup{
    colorlinks = true,
    allcolors = NavyBlue,
    bookmarksnumbered=true,
    pdfpagelabels=false,
    plainpages = false,
}
\usepackage[capitalize]{cleveref}

\graphicspath{{./figures/final}}
\DeclareGraphicsExtensions{.pdf,.png}

\newcommand{\latin}[1]{\textit{#1}}
\newcommand{\pcnt}[1]{\SI{#1}{\percent}}

\definecolor{basal}{RGB}{0, 0, 255}
\definecolor{prism}{RGB}{255, 0, 0}
\definecolor{pyrca}{RGB}{245, 155, 0}

\newcommand{\Ngrain}{\ensuremath{n_\text{grain}}\xspace}
\newcommand{\Ngrid}{\ensuremath{N}\xspace}

\newcommand{\dgrain}{\ensuremath{d_\text{grain}}\xspace}

\newcommand{\crystaxis}[1]{\ensuremath{\langle{#1}\rangle}}

\newcommand{\basalA}{basal \crystaxis{a}\xspace}
\newcommand{\prismA}{prismatic \crystaxis{a}\xspace}
\newcommand{\pyrCA}{pyramidal \crystaxis{c+a}\xspace}

\newcommand{\caratio}{\ensuremath{c/a} ratio}
\newcommand{\slipresistance}[1][]{\ensuremath{\xi^{#1}}}
\newcommand{\slipresistancerate}[1][]{\ensuremath{\dot{\xi}^{#1}}}
\newcommand{\initialresistance}[1][]{\ensuremath{\xi_0^{#1}}}
\newcommand{\finalresistance}[1][]{\ensuremath{\xi_\infty^{#1}}}
\newcommand{\hardeningref}{\ensuremath{h_0}}
\newcommand{\slipsysteminteraction}[1][]{\ensuremath{h_{#1}}}
\newcommand{\hardeningexp}{\ensuremath{a}}
\newcommand{\rss}[1][]{\ensuremath{\tau^{#1}}}
\newcommand{\slipdirection}[1][]{\ensuremath{\vctr{m}^{#1}}}
\newcommand{\slipplane}[1][]{\ensuremath{\vctr{n}^{#1}}}
\newcommand{\shearrate}[1][]{\ensuremath{\dot\gamma^{#1}}}
\newcommand{\shearrateref}{\ensuremath{\shearrate_0}}
\newcommand{\stressexp}{\ensuremath{n}}
\newcommand{\Taylorfactor}{\ensuremath{M}}

\newcommand{\ie}{\latin{i.\hspace{1pt}e.}\xspace}
\newcommand{\eg}{\latin{e.\hspace{1pt}g.}\xspace}

\newcommand{\cf}{\latin{cf.}\xspace}

\newcommand{\term}[1]{#1\xspace}

\newcommand{\PK}{\term{Piola}--\term{Kirchhoff}}

\newcommand{\Mandel}{\term{Mandel}}

\newcommand{\field}[1]{\ensuremath{\mathbb{#1}}}
\newcommand{\tnsr}[1]{\ensuremath{\vb{#1}}}
\newcommand{\vctr}[1]{\ensuremath{\vb{#1}}}

\newcommand{\transpose}[1]{\ensuremath{{#1}^{\text T}}}
\newcommand{\inverse}[1]{\ensuremath{{#1}^{-1}}}

\newcommand{\dblCont}{:}

\newcommand{\identity}{\tnsr I}

\newcommand{\stiffness}{\ensuremath{\field C}}
\newcommand{\F}{\ensuremath{\tnsr F}}
\newcommand{\Fdot}[1][]{\ensuremath{\dot{\tnsr F}_\text{#1}}}
\newcommand{\Fdotcomp}[1]{\ensuremath{\dot F_{#1}}}
\newcommand{\Fp}{\ensuremath{\tnsr F_\text{p}}}
\newcommand{\Fpdot}{\ensuremath{\dot{\tnsr F}_\text{p}}}
\newcommand{\Fi}{\ensuremath{\tnsr F_\text{i}}}
\newcommand{\Fidot}{\ensuremath{\dot{\tnsr F}_\text{i}}}
\newcommand{\Fe}{\ensuremath{\tnsr F_\text{e}}}
\newcommand{\Lp}{\ensuremath{\tnsr L_\text{p}}}
\newcommand{\Li}{\ensuremath{\tnsr L_\text{i}}}

\newcommand{\fPK}[1][]{\ensuremath{{\tnsr P}_\text{#1}}}
\newcommand{\sPK}{\ensuremath{\tnsr S}}
\newcommand{\Mp}{\ensuremath{\tnsr M_\text{p}}}
\newcommand{\Mi}{\ensuremath{\tnsr M_\text{i}}}

\newcommand{\twoonefour}{1 (2:1:4)\xspace}
\newcommand{\onetwofour}{2 (1:2:4)\xspace}
\newcommand{\onefourfour}{3 (1:4:4)\xspace}
\newcommand{\onetwotwo}{4 (1:2:2)\xspace}
\newcommand{\twotwoone}{5 (2:2:1)\xspace}
\newcommand{\oneoneone}{6 (1:1:1)\xspace}
\newcommand{\oneonetwo}{7 (1:1:2)\xspace}
\newcommand{\oneonefour}{8 (1:1:4)\xspace}

\newcommand{\onetwo}{1 (1:2)\xspace}

\newcommand{\oneone}{3 (1:2)\xspace}
\newcommand{\oneponeone}{4 (1.1:1)\xspace}
\newcommand{\twoone}{5 (2:1)\xspace}

\DeclareSIUnit\micron{\micro\metre}

\newcolumntype{M}[1]{>{\centering\arraybackslash}m{#1}}

\newsavebox{\measurebox}

\journal{Acta Materialia}

\begin{document}
\date{August 15, 2023}
\begin{frontmatter}
    \title{A computational study of how surfaces affect slip family activity}

   \author[MSU_PA]{Cathy Bing}
   \ead{zhangrux@msu.edu}

   \author[MSU_CHEMS]{Thomas R. Bieler}
   \ead{bieler@egr.msu.edu}

   \author[MSU_CHEMS]{Philip Eisenlohr\corref{PE}}
   \ead{eisenlohr@egr.msu.edu}

   \cortext[PE]{Corresponding author}

   \affiliation[MSU_PA]{
                   organization={Physics and Astronomy,
                                 Michigan State University},
                   city={East Lansing},
                   state={MI},
                   statesep={},
                   postcode={48824},
                   country={USA}}

   \affiliation[MSU_CHEMS]{
                   organization={Chemical Engineering and Materials Science,
                                 Michigan State University},
                   city={East Lansing},
                   state={MI},
                   statesep={},
                   postcode={48824},
                   country={USA}}

   \begin{abstract}
    Plastic deformation behavior is most conveniently assessed by characterization on a surface, but whether such observations are representative of bulk properties is uncertain.
    Motivated by reported inconsistencies in slip resistance probed at different depths, we investigated (i) whether the average slip family activity is affected by the presence of a surface and (ii) how the kinematic nature of available slip families influences a potential surface effect.
    The slip family activity as a function of distance from the surface was extracted from full-field crystal plasticity simulations of random polycrystalline hexagonal close-packed (HCP) and body-centered cubic (BCC) metals as examples of mixed in contrast to universally-high numbers of slip systems per family.
    Under certain conditions, a deviation from bulk slip activity is observed up to about two grains from the surface.
    For the easiest (least slip-resistant) family, a surface effect of decreasing activity with depth emerges if the number of slip systems falls below about six.
    For harder families, slip activity always increases with depth.
    These phenomena are explained on the basis of varying constraints with depth in connection with the kinematic properties of slip families in the material.
   \end{abstract}
   \end{frontmatter}

\section{Introduction}
\label{sec: Introduction}

The evolution of dislocation content in a plastically deforming grain of a polycrystalline material generally results from a complex interaction between dislocation activity on multiple slip systems.
The evolving dislocation structure results in differing resistance\footnote{frequently termed critical resolved shear stress (CRSS), usually associated with yielding, and sensitive to alloy content and microstructural defects} to further slip on each slip system.
As such, probing the activity of any potential slip system and its associated resistance is a valuable tool to understand the evolution of structure and deformation resistance under load.
Moreover, the quantification of slip activity and deformation resistance is essential to formulate and validate models of crystal plasticity.
Such models, when applied to simulate the heterogeneity of deformation usually exhibited in polycrystalline materials, allow interrogation of, for instance, the (spatially rare) events that are at the root of material degradation and ultimate failure.

The task outlined above becomes even more complex for materials that have multiple slip families, such as hexagonal (HCP) or body-centered cubic (BCC) metals, \eg, Ti and Mg, or Fe.
Over the past few decades, several techniques have been developed to extract essential information for understanding heterogeneous plasticity.
In the following, we briefly summarize the major progress achieved specifically for HCP metals as an exemplar of the challenges.

The most intuitive way is to exclusively activate one selected slip system by intentionally orienting a single crystal relative to the load, as done, for instance, by \citet{10.1007/s11661-002-0153-y}, and confirming the intended activity by surface slip trace observation.
However, if some slip families are significantly harder (\eg \pyrCA in HCP metals), an easier slip family might be activated sooner even when the targeted (harder) system is in the most favored orientation (has the highest Schmid factor), thus frustrating the independent probing of such hard families.

Alternatively, knowledge can be gained from simulating nano-indentation into a single grain using a suitable constitutive model, and fitting its parameters (\eg slip family resistances) to match the experimental response.
Multiple features have been used as the basis for comparison between simulation and experiment, such as the load--displacement response and surface topography \citep{10.1557/jmr.2011.334} or variation of hardness with crystallographic orientation \citep{10.1016/j.actamat.2014.03.014}.
Important challenges for this approach include the uniqueness of the identified parameter set \citep{10.3390/nano12030300} and the reproducibility of the measured response, caused, for instance, by the variability of initial defect content and uneven surface finish \citep{10.1557/jmr.1999.0302, 10.1016/j.ijplas.2008.02.009, 10.1016/j.wear.2009.07.015, 10.1007/s11340-021-00813-7}.

Diffraction-based methods offer a means to directly measure the distortion of the unit cell that can be translated into the stress tensor, and are used with various deconvolution techniques to identify slip resistances.
\citeauthor{10.1016/j.matdes.2021.109543} measured the evolution of lattice plane spacings with increasing plastic strain in Mg by Multireflection Grazing Incidence X-ray Diffraction (MGIXD) \citep{10.1016/j.matdes.2021.109543}, which only probes the surface grains, and by neutron diffraction in bulk grains \citep{10.1016/j.msea.2020.140400}.
The results differed and were fitted with a self-consistent crystal plasticity homogenization model, \ie embedding each of the simulated grains in an infinite material of their average strength.
The slip family resistances required to match the measurements exhibited a larger difference among slip resistances (a wider spread) and a lower \basalA resistance for the surface than the bulk \citep[table 3]{10.1016/j.matdes.2021.109543}, suggesting a systematic difference in deformation activity for different slip families between surface and bulk grains.

Furthermore, synchrotron-based high-energy X-ray diffraction microscopy with far-field detection (ff-HEDM) is able to measure the average deviatoric stress tensor and lattice orientation of most grains within an illuminated sample volume of typically \SI{1}{\cubic\milli\meter} at multiple strain levels before and after plastic yielding.
From such data, the shear stress on every slip system in each grain was derived and the evolving maximum values per slip family were interpreted as the characteristic slip resistances, \eg, \citep{10.1016/j.scriptamat.2017.08.029,10.1016/j.actamat.2021.117372}.
Similar to the concern raised above in the case of oriented single crystals, the highest apparent slip resistances might not correlate with actual activity on those slip systems, but be a (trivial) consequence of plastic flow carried by other (easier) slip systems, thus limiting the overall maximum stress attained in such grains.
One possibility to circumvent this uncertainty regarding actual slip activity is to compare the measured grain lattice orientation change to that expected for the assumed slip activity \citep{10.1016/j.actamat.2021.117372}.
Another approach pursued by \citet{10.1016/j.actamat.2017.05.015, 10.1016/j.actamat.2018.05.065} is based on the insight that dominant \prismA slip causes a lattice rotation about the \crystaxis{c}-axis and dominant \basalA slip about an axis perpendicular to \crystaxis{c}.
For grains that exhibited one of these two special rotations, the maximum resolved shear stress for the corresponding slip family was then averaged and interpreted as the respective slip resistance in commercially-pure Ti as sample material.

In contrast, a relatively inexpensive and straightforward approach to directly obtain the ratios between slip family resistances was proposed by \citet{10.1016/j.actamat.2013.08.042} based on the ratios of expected and observed frequencies of surface slip trace observations for each slip family and used, for instance, to investigate the influence of alloying elements on slip resistances in Mg \citep{10.1016/j.jma.2021.03.005}.

\begin{figure}
    \centering
    \includegraphics{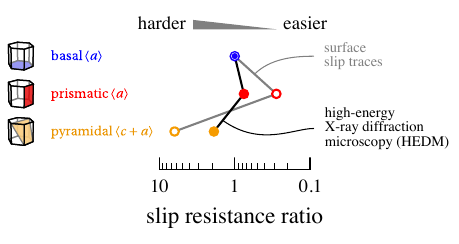}
    \caption{Slip resistance of (first-order) \pyrCA (orange) and \prismA (red) relative to that of the \basalA (blue) slip family as reported in two investigations sharing identical sample material of commercially pure Ti but differing depths of probed volume \citep{10.1016/j.actamat.2013.08.042, 10.1016/j.actamat.2017.05.015} (open and filled circles, respectively).
    The surface slip resistances (open circles) exhibit a wider spread.
    }
    \label{fig: 2methods}
\end{figure}

In both cases where the same material was investigated with techniques that probe either (near-)surface or bulk locations, \ie Mg \citep{10.1016/j.actamat.2019.11.023, 10.1016/j.matdes.2021.109543} and commercially-pure Ti \citep{10.1016/j.actamat.2013.08.042, 10.1016/j.actamat.2017.05.015} (see also (\cref{fig: 2methods}), the spread in identified slip resistances turned out to be systematically wider when extracted close to the surface.
As the idea of systematically different intrinsic slip resistances at different material depths seems unlikely, \citet{10.1016/j.actamat.2019.11.023} conjectured that a universal surface influence may exist because the mechanical constraint that opposes plastic deformation of a grain within a polycrystal gradually decreases with proximity to the surface.      
The open question addressed in this work is when and how surfaces affect the slip family activity observed in a grain and whether the magnitude of this effect can be quantified and predicted such that inexpensive surface measurements can be used to extract bulk slip resistances.

To address the above question, we set up crystal plasticity simulations of thick polycrystalline film as well as bulk polycrystals as a reference.
To investigate the influence of dissimilarity among slip families, HCP and BCC metals were used as examples for mixed in contrast to universally-high slip family multiplicity.%
\footnote{The number of distinct slip systems in a family is called its ``multiplicity''.}
In HCP, \basalA, \prismA, and (first-order) \pyrCA were considered, whereas \hkl{110}\hkl<111> and \hkl{112}\hkl<111> slip families were included for BCC.

\section{Methods}
\label{sec: Methods}

The simulations used the grid solver of the Düsseldorf Advanced Material Simulation Kit (DAMASK, \citep{10.1016/j.commatsci.2018.04.030}), which employs a numerically efficient spectral method  \citep{10.1016/j.ijplas.2014.02.006, 10.1007/978-981-10-6884-3_80} to solve the mechanical boundary value problem of static equilibrium.
DAMASK uses a finite-strain framework in which the total deformation gradient
\begin{align}
    \F &= \pdv{\vctr y}{\vctr x}
\end{align}
maps points \vctr x from the undeformed configuration to points \vctr y in the deformed configuration.
The deformation gradient is decomposed as
\begin{align}
    \F &= \Fe \Fi \Fp,
\end{align}
where \Fp\ is a lattice-preserving, inelastic deformation gradient that maps to the plastic configuration, \Fi\ is a lattice-distorting, inelastic deformation gradient, \eg\ thermal expansion or crack opening, mapping further to the eigenstrain configuration, and \Fe\ is an elastic deformation gradient that maps from the eigenstrain to the deformed configuration.

\subsection{Material constitutive description}
\label{sec: constitutive}

The triple decomposition, \ie consideration of \Fi, is beneficial when mimicking the effect of a surface through the introduction of a slice of dilatational material within a periodic simulation environment.

The two inelastic deformation gradients evolve at rates
\begin{subequations}
\begin{align}
    \Fpdot &= \Lp \Fp
    \\
    \Fidot &= \Li \Fi
\end{align}
\end{subequations}
governed by the velocity gradients \Lp\ and \Li\ that are driven by their respective work-conjugate stress measures, \ie, the \Mandel stresses
\begin{subequations}
\begin{align}
    \Mp &= \transpose{\Fi} \, \Fi \, \sPK
    \\
    \Mi &= \det{\inverse{\Fi}} \, \Fi \, \sPK \, \transpose{\Fi}
\end{align}
\end{subequations}
Both Mandel stresses are mappings of the second \PK stress
\begin{align}
    \sPK &= \stiffness\dblCont \frac{1}{2} \transpose{\Fi}\left(\transpose{\Fe}\,\Fe-\identity\right)\Fi,
\end{align}
which follows Hooke's law, where \stiffness\ is the fourth-order elastic tensor.

The plastic velocity gradient
\begin{align}
    \Lp = \shearrate[\alpha] \slipdirection[\alpha] \otimes \slipplane[\alpha]
\end{align}
adds shear contributions from all considered slip systems (indexed by \(\alpha\) and implicitly summed over repeated indices) with shear rate \shearrate\ and unit vectors \slipdirection\ along the slip direction and \slipplane\ along the slip plane normal.
The resolved stress on each slip system follows as the projection
\begin{align}
    \rss[\alpha] &= \Mp \dblCont \left(\slipdirection[\alpha] \otimes \slipplane[\alpha]\right)
\end{align}
and generally drives the shear deformation at a rate
\begin{align}
    \shearrate[\alpha]=
    \shearrateref
    \left|\frac{\rss[\alpha]}{\slipresistance[\alpha]}\right|^{\stressexp}
    \operatorname{sgn}\left(\rss[\alpha]\right),
\end{align}
where \shearrateref\ is a reference shear rate and \stressexp\ denotes the stress exponent.

The resistance \slipresistance[\alpha]\ to crystallographic slip along each slip system is modeled according to the phenomenological constitutive description introduced by \citet{10.1016/0001-6160(82)90005-0}.
Following \citet{10.1016/0749-6419(89)90025-9}, the asymptotic evolution of each slip resistance from an initial value \initialresistance[\alpha]\ to a saturation value \finalresistance[\alpha] has contributions due to slip on all operating systems:
\begin{align}
    \slipresistancerate[\alpha] &=
    \hardeningref
    \abs{1-\frac{\slipresistance[\alpha]}{\finalresistance[\alpha]}}^\hardeningexp
    \operatorname{sgn}\left(1-\frac{\slipresistance[\alpha]}{\finalresistance[\alpha]}\right)
    \slipsysteminteraction[\alpha\beta]
    \abs{\shearrate[\beta]},
    \label{eq: slip resistance}
\end{align}
where \hardeningref\ is a reference hardening parameter, \hardeningexp\ is the hardening exponent, and \slipsysteminteraction[\alpha\beta] characterizes the slip system interactions.

A similar constitutive description is used to describe the evolution of \Fi\ in the case of a ``virtual air'' layer introduced in \cref{sec: Geometry} (see \citep{10.1016/j.scriptamat.2017.09.047, 10.1016/j.commatsci.2018.04.030} for details).

Parameters used in the simulations are listed in \cref{tbl: parameters}.

\begin{table}
    \centering
    \caption{Constitutive parameters of exemplary HCP and BCC materials and the virtual air.}
    \begin{tabular}{l r r r c}
         \toprule
          \textbf{Property} & \multicolumn{3}{c}{\textbf{Value}} & \textbf{Unit}\\
         \cmidrule(lr){2-4}
          & HCP & BCC & Air & \\
         \midrule
          \caratio & 1.587 & \(\cdot\) & \(\cdot\) & \\
          \initialresistance & various & various & 0.03 & \si{\mega\pascal} \\
          \(\finalresistance/\initialresistance\) & 3 & 3 & 2 & \\
          \hardeningref & 0.2 & 1 & \num{e-3} & \si{\giga\pascal} \\
          \multirow{2}{*}{\slipsysteminteraction[\alpha\beta]} & \multirow{2}{*}{1} & 1 (coplanar) & \multirow{2}{*}{\(\cdot\)} & \\
           & & 1.4 (non-coplanar) & & \\
          \shearrateref & \num{e-3} & \num{e-3} & \num{e-3} & \si{\per\s} \\
          \stressexp & 20 & 20 & 5 & \\
          \Taylorfactor & \(\cdot\) & \(\cdot\) & 3 & \\
          \hardeningexp & 2 & 2 & 2 & \\

         \bottomrule
    \end{tabular}
    \label{tbl: parameters}
\end{table}

\subsection{Geometry}
\label{sec: Geometry}

\begin{figure}[b]
    \centering
        \includegraphics[width=\linewidth]{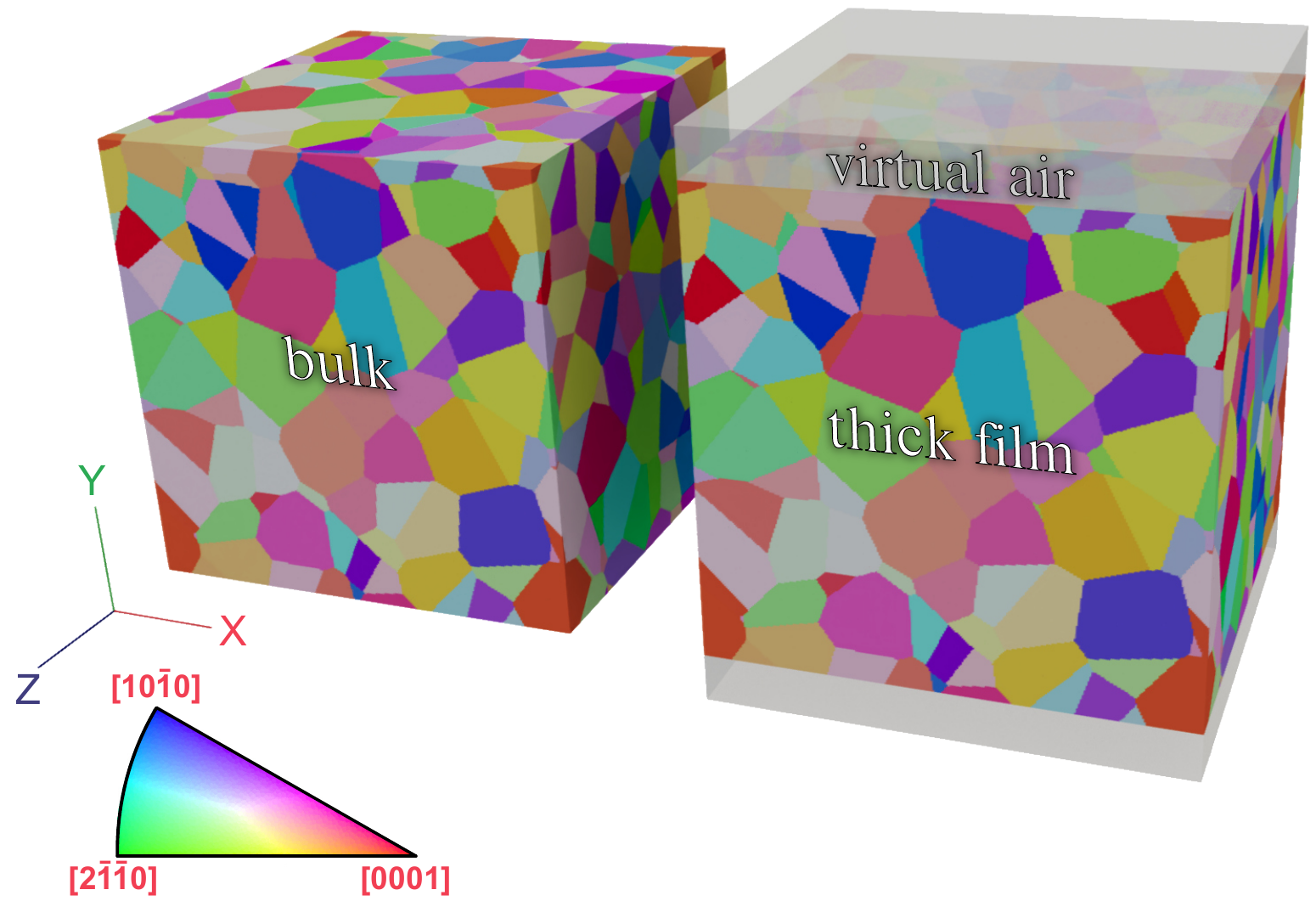}
\caption{
        Exemplary unit cells of
        (left) periodic Voronoi tessellation with 400 grains mimicking a polycrystalline bulk structure, and
        (right) the same polycrystalline bulk structure but sandwiched by a layer of dilatational, soft, and compliant material (``virtual air'', translucent) to introduce two, essentially free, surfaces on top and bottom.
        Grain color reflects crystallographic direction along the loading axis \(x\) (with inverse pole figure coloring of hexagonal symmetry).
        }
\label{fig: structure examples}
\end{figure}

To investigate how surfaces influence the slip family activity, polycrystalline structures with and without the presence of surfaces were constructed (\cref{fig: structure examples}) and labeled as ``bulk'' and ``film'', respectively.

A periodic polycrystalline bulk structure (\cref{fig: structure examples} left) contains \(\Ngrain = 400\) randomly oriented grains resulting from a Voronoi tessellation of a random Poisson point distribution within a cubic volume discretized by \(\Ngrid = 96 \times 96 \times 96\) equidistant grid points.
Reported values of slip family activity represent cumulative shear arising from each of the considered slip families averaged over the whole volume.

A freestanding polycrystalline film that is multiple grains thick results from inserting a layer of dilatational, low-strength, low-stiffness material (``virtual air'') into a periodic polycrystalline bulk structure to mechanically decouple the top and bottom faces (\cref{fig: structure examples} right).
Since the layer of virtual air can only exert a minuscule normal force (along \(y\)), both interfaces effectively act like free surfaces.
In the case of the film structure, values of slip family activity are averaged per slice normal to the surface and reported as a function of distance to the nearest surface, \ie as a depth profile.

\begin{figure*}
    \centering
        \includegraphics{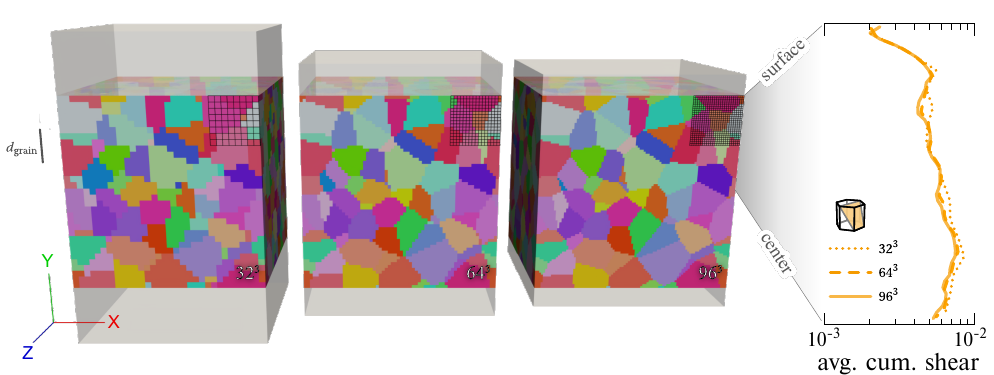}
    \caption{
        The same film structure discretized by \(\Ngrid = 32^3\), \(64^3\), and \(96^3\) voxels within the metal domain, indicated by the black wireframe.
        The translucent air layer is \(10+10\) voxels thick in all cases.
        The film center is about 3 grain diameters from the surface.
        The depth profile on the right shows the average cumulative \pyrCA shear in each slice after \pcnt{5} strain along \(x\) for the three different grid resolutions.
        }
    \label{fig: convergence}
\end{figure*}

Grid convergence was tested with three resolutions for the film structure as shown in \cref{fig: convergence} (left) while keeping the number of grid points discretizing the air layer along the surface normal constant.
Since the chosen constitutive law is scale-independent, the coordinate system (specifically the distance to the surface) is normalized by the average grain diameter \dgrain%
\footnote{\(\Ngrid/\Ngrain = \frac{4\pi}{3} \left(\frac{\dgrain}{2}\right)^3\)}
as the only scale-determining quantity.
\Cref{fig: convergence} (right) shows the slip activity depth profile observed for \pyrCA, which is the worst converging one of all slip families in this example.
Its slip activity is virtually identical between grid resolutions of \(64^3\) and \(96^3\) in the metal domain.
Hence, \(N=96^3\) was considered as a suitably converged resolution.
Note that the first point from the surface in \cref{fig: convergence} (first layer of non-air voxels) is deviating from this overall converging behavior, most likely caused by the Gibbs phenomenon in response to the sudden property contrast across this interface.

\subsection{Boundary conditions}
\label{sec: Boundary conditions}

Owing to the numerical solution strategy, all simulations obey periodic boundary conditions such that only volume-averaged stress or deformation values can be prescribed.
Both the ``bulk'' and ``film'' structure are deformed along the \(x\) direction under mixed boundary conditions that are targeting a unidirectional stress response.
Specifically, the volume-averaged deformation gradient rate \Fdot[] and work-conjugate first \PK\ stress \fPK[] values
\begin{align}
    \Fdot[bulk] &=
    \begin{bmatrix}
    \num{e-3} & 0 & 0 \\
    0 & \cdot & 0 \\
    0 & 0 & \cdot \\
    \end{bmatrix} \unit{\per\second}\quad\text{and}
    &
    \fPK[bulk] &=
    \begin{bmatrix}
    \cdot & \cdot & \cdot \\
    \cdot & 0 & \cdot \\
    \cdot & \cdot & 0 \\
    \end{bmatrix} \si{\pascal}
    \label{eq: boundary conditions bulk}
\\[\baselineskip]
    \Fdot[film] &=
    \begin{bmatrix}
    \num{e-3} & 0 & 0 \\
    0 & 0 & 0 \\
    0 & 0 & \cdot \\
    \end{bmatrix} \unit{\per\second}\quad\text{and}
    &
    \fPK[film] &=
    \begin{bmatrix}
    \cdot & \cdot & \cdot \\
    \cdot & \cdot & \cdot \\
    \cdot & \cdot & 0 \\
    \end{bmatrix} \si{\pascal}
    \label{eq: boundary conditions film}
\end{align}
were prescribed for a duration of \SI{10}{\second} and \SI{50}{\second} resulting in a final extension of \pcnt{1} and \pcnt{5}, respectively, as it is helpful to include the influence of work hardening in the scope of this work.
A dot `\(\cdot\)' in \cref{eq: boundary conditions bulk,eq: boundary conditions film} indicates that the complementary condition is prescribed.
Rather than adopting the bulk boundary condition for the film structure, the numerical convergence of the film improved notably (with an insignificant effect on the result) when preventing any extension along the surface normal, \ie \(\Fdotcomp{yy} = 0\), such that any contraction of the crystalline volume along \(y\) is accommodated by a corresponding extension in the thickness of the (essentially stress-free) layer of virtual air.

\subsection{Parametric study}
\label{subsec: studied parameters}

To understand the effects of (i) crystal elasticity, (ii) differences in slip resistance per family, and (iii) multiplicity in easy and hard slip families on a potential surface effect, a parametric study along these three dimensions was performed.

\subsubsection{Elasticity}
\label{subsubsec: Elasticity}

\begin{figure}
    \centering
    \includegraphics[width=\linewidth]{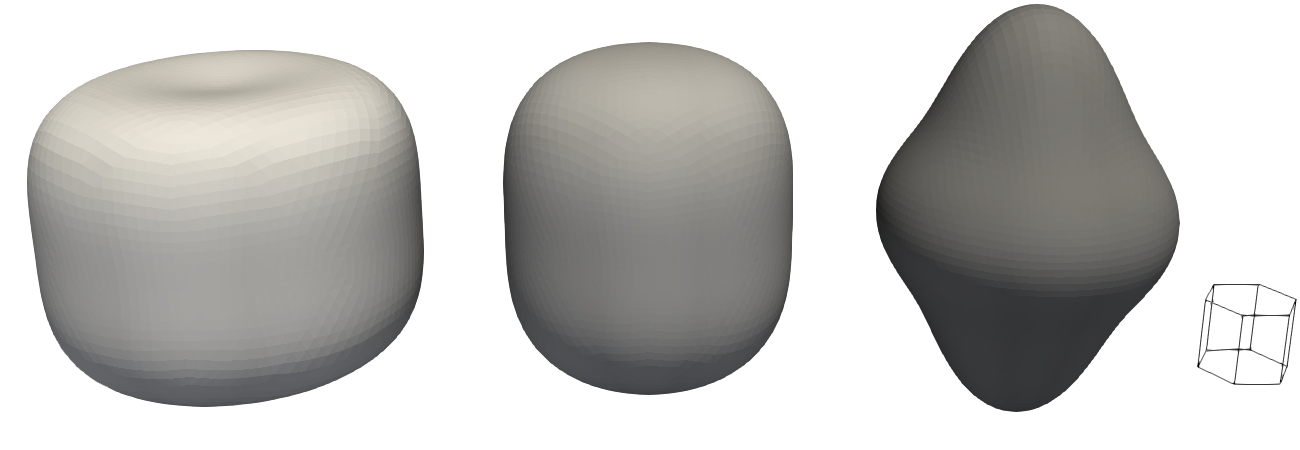}
    \caption{
    Directional stiffness in lattice frame coordinates illustrating each of the three representative instances of HCP elastic anisotropy listed in \cref{tbl: elastic constants}; from left to right and light to dark: ``marshmallow'', ``capsule'', and ``spin''.
    }
    \label{fig: elasticity}
\end{figure}

\begin{table}
    \centering
    \caption{Elastic stiffness tensor components of (isotropic) virtual air, three sample materials that span the anisotropy range of HCP metals, and an isotropic BCC material.}
    \begin{tabular}{l *{5}{r}}
        \toprule
        \textbf{Material} & \multicolumn{5}{c}{\textbf{Elastic stiffness tensor} / \unit{\giga\pascal}}
        \\ \cmidrule(lr){2-6}
          & \(C_{11}\) & \(C_{12}\) & \(C_{13}\) & \(C_{33}\) & \(C_{44}\)
        \\ \midrule
        virtual air & \num{e-1} & \num{e-3} & \(\cdot\) & \(\cdot\) & \(\cdot\)
        \\ \midrule
         HCP
        \\ \cmidrule(lr){1-1}
         marshmallow (Cd, Zn) & 165 & 5 & 55 & 140 & 60
        \\
         capsule (Be) & 110 & 10 & 5 & 130 & 70
        \\
         spin (Co, Mg, Zr) & 165 & 90 & 60 & 190 & 40
        \\ \midrule
        BCC & 270 & 110 & \(\cdot\) & \(\cdot\) & 80
        \\
        \bottomrule
    \end{tabular}
    \label{tbl: elastic constants}
\end{table}

To investigate the influence of elastic anisotropy, HCP was chosen as an example.
Upon inspection of the elemental HCP metals, it turns out that their elastic anisotropy can be clustered into three groups as illustrated in \cref{fig: elasticity}.
\Cref{tbl: elastic constants} summarizes the stiffness tensor components adopted in this study for HCP metals to represent these three distinct shapes (scaled to approximately equal overall magnitude) with some examples in each of the three groups, as well as for BCC metals and virtual air elasticity.

\subsubsection{Slip resistance}
\label{subsubsec: slip resistance}

To discern whether the crystallography or the relative ease/difficulty of activation is the primary cause for a potential surface effect on the slip family activity, several choices of relative slip family resistances provide a comprehensive comparison (see left column of \cref{fig: hcp} and \cref{fig: bcc}).

\subsubsection{Slip family multiplicity}
\label{subsubsec: multiplicity}

To understand whether the slip family multiplicity plays a role in a surface effect, the number of available slip systems in the \pyrCA family was artificially reduced to being equal to \basalA and \prismA, \ie from 12 to 3 for fixed slip resistances and elastic anisotropy (see left column of \cref{fig: multiplicity}).

\section{Results and Discussion}
\label{sec: Results and Discussion}

All parametric studies of film or bulk structures average 50 or 5 independent realizations, containing a total of \(50 \times 400 = \num{20000}\) or \(5 \times 400 = \num{2000}\) grains of random shape and orientation.
\Cref{sec: HCP} considers HCP metals as an example of lattices with mixed multiplicity, whereas BCC metals are investigated in \cref{sec: BCC} as example for universally high multiplicity.
After summarizing the surface effect behavior in \cref{sec: behaviors}, corresponding mechanisms are discussed in \cref{sec: mechanisms}.

\subsection{HCP}
\label{sec: HCP}

\Cref{fig: hcp} displays the HCP slip family activity along the depth of the film structures (vertical curves) and in the bulk polycrystal (filled squares below bottom scale) up to \pcnt{1} and \pcnt{5} strain with their corresponding slip resistance ratios shown in the left column.
Data is colored by slip family and the different shades indicate the three elastic anisotropy cases.
Horizontal bars give the central \pcnt{68} of each slip family activity population.
The vertical bar represents the average grain size \dgrain, \ie, the overall distance from the surface to the center of the film structure is about three grains.

\subsubsection{Influence of elastic anisotropy}

The influence of elastic anisotropy diminishes with increasing strain, as reflected by the differently shaded square symbols and associated depth profiles becoming overlaid in the right column of \cref{fig: hcp}.
The variability of slip activity (horizontal bars) with depth is only illustrated at three exemplary depths across the three elastic anisotropy instances as it turned out to not strongly depend on the specific anisotropy.
Overall, the elastic anisotropy variations do not affect the shape of the depth profiles.

\subsubsection{Bulk slip family activity}

Naturally, the slip systems with the lowest slip resistance (rightmost in first column with reversed scale) values show the greatest activity.
The (relative) spread in the slip family activity directly reflects the imposed slip resistance ratios.
For instance, the larger ratios in row \oneonefour result in wider activity spread than the smaller ratios in row \oneonetwo.
Thus, whenever the slip resistance ratios narrow in response to strain hardening (filled circles compared to open ones), less favored slip families become relatively more active (shrinking distance between squares in column 3 compared to column 2).

\subsubsection{Depth profiles of slip family activity}

\begin{figure*}
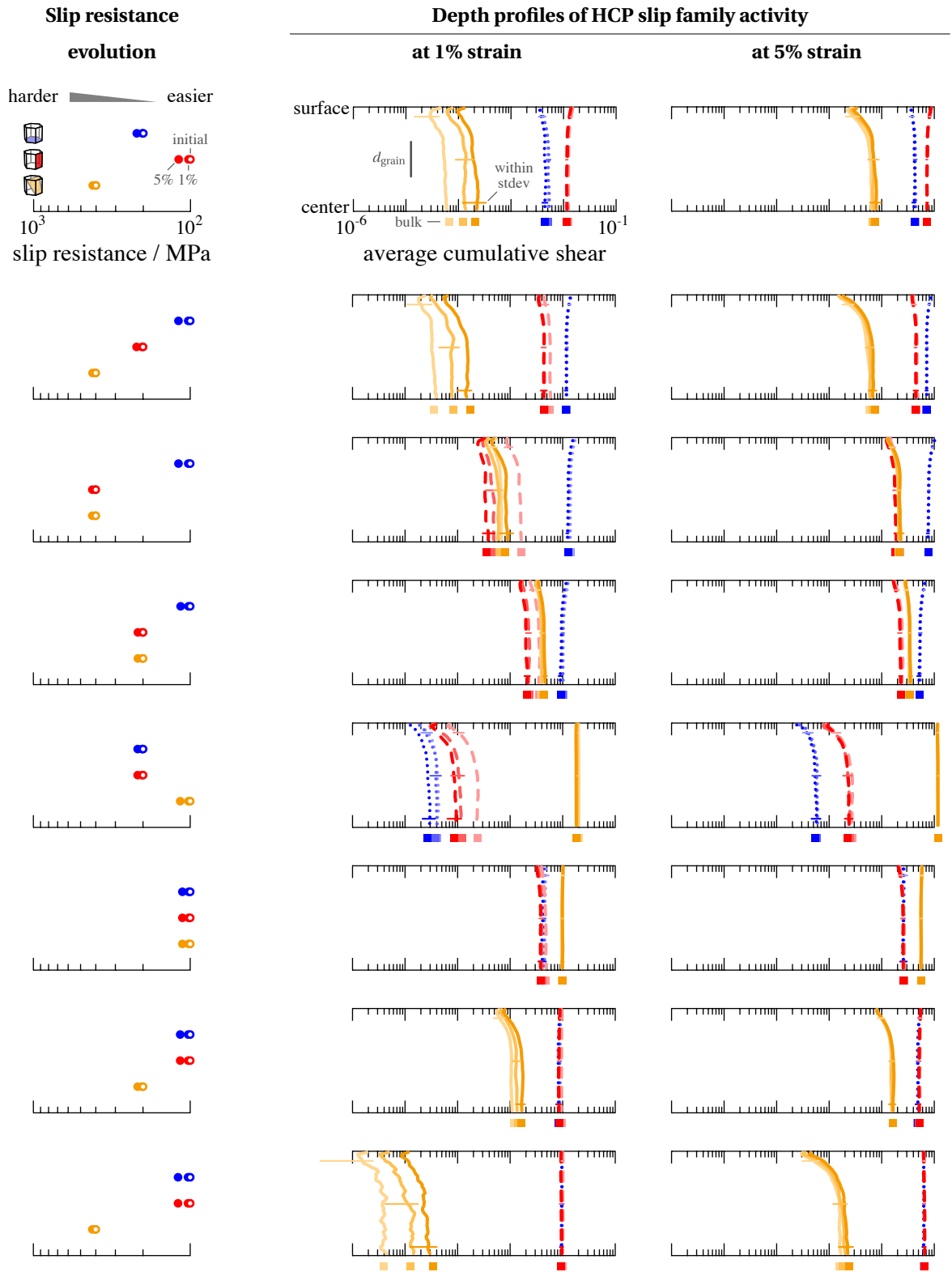

    \centering
    \begin{tabularx}{0.9\linewidth}{>{\centering\arraybackslash}Xcc}
        \textbf{Slip resistance}&\multicolumn{2}{c}{\textbf{Depth profiles of HCP slip family activity}} \\
        \cmidrule(lr){2-3}
        \textbf{evolution}&\textbf{at 1\xspace{}\% strain}&\textbf{at 5\xspace{}\% strain}
        \\[3mm]
        \xintFor #1 in {7, 3, 6, 2, 0, 1, 4, 5}
        \do {%
            \includegraphics{xi_#1_wh}
            &\includegraphics{#1_inc10}
            &\includegraphics{#1_inc50}
        \\[2mm]
        }
    \end{tabularx}
    \caption{
        For eight cases of different slip resistance ratios in the left column, the right columns present the resulting depth profiles of slip family activity (vertical curves) at two strain levels.
        Closed circles in the left column correspond to the 95th percentile of slip resistances, which evolve essentially identically in film and bulk structures for any elastic anisotropy.
        Slip families are represented by different colors (and line styles) with shades reflecting the elastic anisotropy (see \cref{fig: elasticity}).
        The vertical scale in the right columns spans from the film surface to the center, about three grains deep.
        The slip activity profile converges below about one to two grain diameters to the corresponding bulk activity, which is represented as filled squares below the bottom scale.
        Horizontal bars exemplify the (virtually depth-independent) spread of the central \pcnt{68} of each slip family activity population at three separate depths.
        }
    \label{fig: hcp}
\end{figure*}


At a depth exceeding about \(2\,\dgrain\), the activity in the film is essentially the same as that observed in the bulk, \ie the value in the lower third of each curve is constant and equal to the corresponding square indicating bulk slip activity.
In contrast, when approaching the surface (within \(2\,\dgrain\)), the slip activity of a film usually deviates from the bulk response.
We note that the distance of about two average grain diameters beyond which a surface effect fades out is consistent with the results reported by \citet{10.1007/s11012-015-0281-2} in their investigation of ``columnarity'', \ie, how far away a change in grain structure is influencing the mechanical response in a bulk polycrystal.%
\footnote{We note the slight conceptual difference of ``structural change'' being the \emph{removal} of grains beyond the surface in the present study in contrast to an \emph{alteration} of those grains in ref.~\citep{10.1007/s11012-015-0281-2}.}

From the depth profiles presented in \cref{fig: hcp}, two main observations can be made.
The slip activity toward the surface
\begin{enumerate}
    \item decreases for any slip family that is harder than the easiest
    \item increases for the easiest slip family in case of \basalA (row \onetwofour, \onefourfour, and \onetwotwo) and \prismA (row \twoonefour) but not for \pyrCA (row \twotwoone) or combined \basalA and \prismA (last three rows)
\end{enumerate}
The latter observation suggests that the number of available slip systems in the easiest family (or a combination thereof) is a governing factor in the emergence of a surface effect.

\subsubsection{Slip family multiplicity}

\begin{figure}
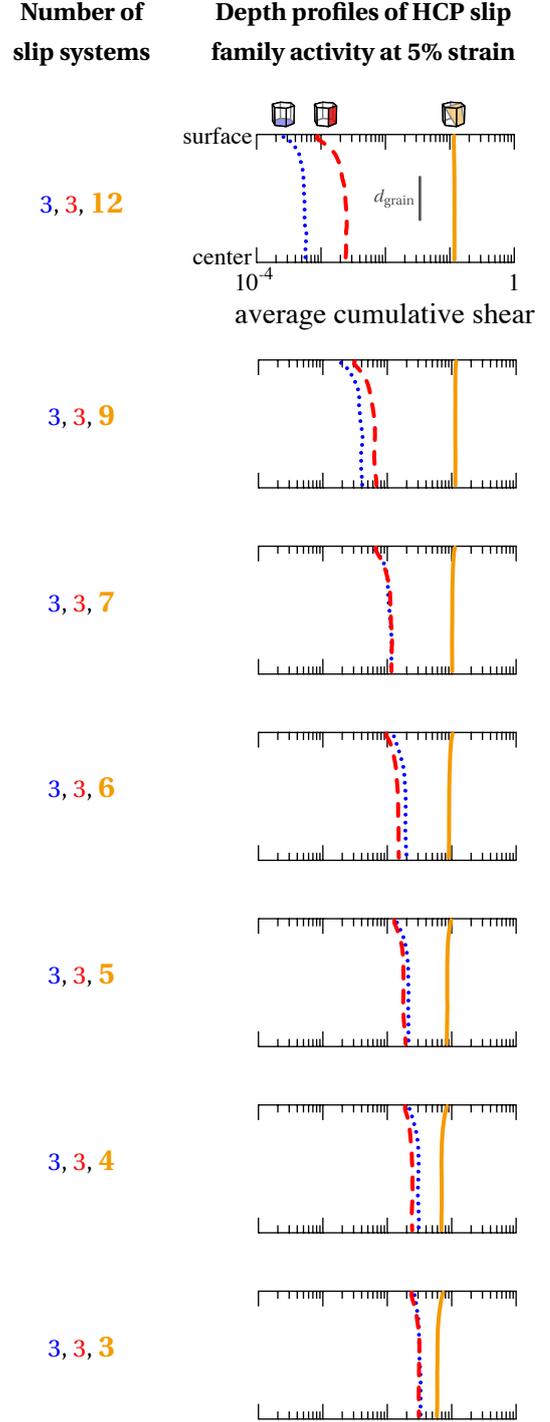

    \centering
    \begin{tabular}{cr}
        \textbf{Number of}
        & \multicolumn{1}{c}{\setlength{\fboxrule}{0pt}\framebox{\textbf{Depth profiles of HCP slip}}}
        \\
        \textbf{slip systems}
        & \multicolumn{1}{c}{\setlength{\fboxrule}{0pt}\framebox{\textbf{family activity at 5\xspace{}\% strain}}}
        \\[3mm]
        \xintFor #1 in {12, 9, 7, 6, 5, 4, 3}
        \do {%
        \textcolor{basal}{3}, \textcolor{prism}{3}, \textbf{\large\textcolor{pyrca}{#1}}
        &\includegraphics[align=m]{00_33#1}
        \\[2mm]
        }
    \end{tabular}
    \caption{
        Slip activity of the three HCP slip families (\basalA, \prismA, \pyrCA) as a function of distance from the surface for a fixed slip resistance ratio (same as row \twotwoone in \cref{fig: hcp}) and elastic anisotropy (``capsule'' in \cref{tbl: elastic constants}) but with a progressively reduced number of \pyrCA slip systems in the family (\ie reduced multiplicity).
        }
    \label{fig: multiplicity}
\end{figure}

To elucidate the influence of the number of available slip systems in the emergence of a surface effect for the easiest slip family, we select row \twotwoone of \cref{fig: hcp} as a test case and reduce the multiplicity of the (here easiest) \pyrCA slip family from 12 to 3.%
\footnote{The ``capsule'' elastic tensor from \cref{tbl: elastic constants} was used.}
\Cref{fig: multiplicity} shows that this change results in a gradual increase of the activity of the two harder families (\basalA and \prismA) but does not qualitatively alter their surface effect, \ie, the blue and red curves remain bent to the left.
In contrast, the \pyrCA family develops an increasingly notable surface effect below a multiplicity of seven.

Such a surface effect is always present if the easiest slip family has low multiplicity.
For instance, the surface effect of \basalA (multiplicity of three) as the easiest family in rows \onetwofour, \onefourfour, and \onetwotwo of \cref{fig: hcp} is very comparable to that of the reduced multiplicity \pyrCA in the bottom row of \cref{fig: multiplicity}.
Similarly, the combination of \basalA and \prismA as easiest families in row \oneonetwo in \cref{fig: hcp} resembles the result for \pyrCA with the same (3+3) multiplicity of six in \cref{fig: multiplicity}.

\subsection{BCC}
\label{sec: BCC}

\begin{figure}
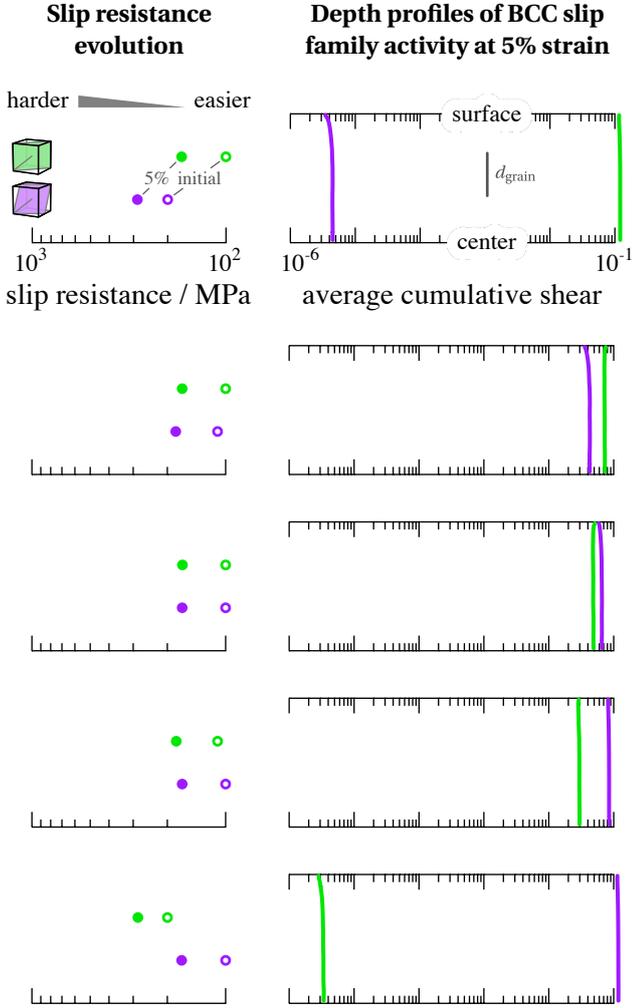

    \centering
    \begin{tabular}{cc}
    \textbf{Slip resistance}&\textbf{Depth profiles of BCC slip} \\
    \textbf{evolution}&\textbf{family activity at 5\xspace{}\% strain}
    \\[3mm]
    \xintFor #1 in {2, 12, 1, 11, 5}
    \do {%
        \includegraphics{xi_#1_wh_1e9}
        &\includegraphics{#1_12_12_1e9}
    \\[2mm]
    }
    \end{tabular}
    \caption{
    For five different slip resistance ratios in the left column, the resulting depth profiles of BCC slip family activity are presented in the right column after \pcnt{5} strain.
    Closed circles in the left column correspond to the 95th percentile of slip resistances evolved after \pcnt{5} strain from initial values (open circles).
    The vertical scale in the right column spans from the film surface to the center, about three grains deep.
    Variation across the central \pcnt{68} of each slip family activity population is small and less than the curve widths.
    The slightly higher activity of \hkl{112} compared to \hkl{110} slip at equal slip resistance is connected to the slightly higher chance for a \hkl{112}\hkl<111> slip system to have the largest Schmid factor.
    }
    \label{fig: bcc}
\end{figure}

The combinations of multiplicity and slip resistance investigated so far suggest that in a situation of high multiplicity for soft as well as hard families, the surface effect is comparable to the results shown in rows \twotwoone and \oneonetwo of \cref{fig: hcp}.
To test this hypothesis, we investigated an exemplary BCC material with two slip families, each having a multiplicity of twelve.
\Cref{fig: bcc} follows the format of \cref{fig: hcp} and compares the slip family depth profiles after \pcnt{5} strain across various slip resistance ratios between the \hkl{110} (green) and \hkl{112} (purple) slip families.

From top to bottom, the observed slip activity follows the changing ratios from strongly favoring \hkl{110} to strongly favoring \hkl{112}.
The decrease of activity with increasing relative slip resistance is asymmetric: \hkl{110} slip (green curves top to bottom) shows a greater decrease than \hkl{112} slip (purple curves bottom to top) at comparable slip resistances.
This is most apparent in the middle row of \cref{fig: bcc}, \ie \hkl{112} is more active than \hkl{110} slip at equal slip resistance.
We note in passing that for common BCC metals, the slip resistance of \hkl{110} and \hkl{112} is similar, perhaps within \pcnt{10} of difference, \ie close to the conditions shown in the middle three rows of \cref{fig: bcc}.
The difference in slip activity despite equally slip-resistant families (about \pcnt{25} in the bulk of row \oneone) can be rationalized by the fact that for a random stress state, the chance that a \hkl{112} slip system experiences the largest resolved shear stress is a few percent higher than for a \hkl{110} slip system.

Compared to what was observed for HCP metals under most conditions, the extent of surface effect in BCC cases is subtle to non-existent.

\subsection{Types of surface effect manifestations}
\label{sec: behaviors}

Overall, three distinct patterns \ref{behavior: hard}, \ref{behavior: soft incomplete}, and \ref{behavior: soft complete} are observed for the surface effect on slip family activity:
\begin{enumerate}[label=\alph*)]
    \item \label{behavior: hard}
    The activity of slip families that are harder than the easiest one is \emph{decreased} near the surface relative to the interior, \eg \prismA and \pyrCA in row \onetwofour of \cref{fig: hcp}, or \basalA and \prismA in row \twotwoone of the same figure.
    \item \label{behavior: soft incomplete}
    Conversely, the surface activity is \emph{increased} for the easiest slip family provided its multiplicity is low, such as \basalA in rows \onetwofour of \cref{fig: hcp}, or \pyrCA in the bottom rows of \cref{fig: multiplicity}.
    \item \label{behavior: soft complete}
    If the easiest slip family has considerable multiplicity, the surface effect on its activity is \emph{virtually unnoticeable}, as demonstrated by the essentially straight curves of, for instance, \pyrCA in top rows of \cref{fig: multiplicity}, combined \basalA and \prismA (featuring six slip systems in total) in row \oneonetwo and \oneonefour of \cref{fig: hcp}, or \hkl{112} slip in row \oneponeone of \cref{fig: bcc}.

\end{enumerate}

Moreover, the surface effect intensifies with
\begin{enumerate}
    \item larger slip resistance contrast, \eg all three slip families in row \onefourfour of \cref{fig: hcp} show a larger curvature than in row \onetwotwo, a more drastic surface effect on \pyrCA slip activity is observed in row \oneonefour compared to row \oneonetwo, and for \hkl{110} in row \twoone compared to row \oneponeone of \cref{fig: bcc};

    \item larger multiplicity contrast, \eg, with a slip resistance ratio of 2, \cref{fig: hcp} shows more pronounced surface effects than \cref{fig: bcc}.
\end{enumerate}

\subsection{Mechanisms}
\label{sec: mechanisms}

Statistically, the best-aligned slip systems of the easiest slip family are activated first under loading, and the shape change caused by their activity generally leads to incompatibility with neighboring grains.
In consequence, the stress state changes progressively favoring slip systems that help maintain compatibility.
The overall compatibility constraints are naturally relaxed closer to the surface than further into the bulk.
These ideas are at the core of the following explanations of the three different manifestations of the surface effect.

If the easiest slip family has numerous members (high multiplicity, typically more than six slip systems), chances are high that tighter compatibility requirements with increasing depth can be fulfilled by activating additional members of this (easiest) family.
Therefore, if the observed slip activity is aggregated at the \emph{family} level, there will be no substantial influence of changing compatibility constraints with depth on the overall activity of the easiest family, leading to pattern \ref{behavior: soft complete}, \ie, no surface effect for the easiest family.

However, if the multiplicity of the easiest family is low, an increase in mechanical constraint, \ie toward the interior, will decrease the chance to maintain compatibility by slip activity of that easiest family \emph{alone}, resulting in decreasing activity of the easiest family with increasing depth as reflected in pattern \ref{behavior: soft incomplete}.

Because harder families are generally activated in response to mechanical constraint, their activity will decrease towards the surface, giving rise to pattern \ref{behavior: hard}, independent of whether or not softer families exhibit a surface effect.

When two or more low-multiplicity families share similar slip resistance values, this effectively increases the overall multiplicity of the resulting composite family and generally decreases a surface effect compared to its separate members.
Examples are \basalA and \prismA in row \oneonetwo and \oneonefour of \cref{fig: hcp}, which, combined, act in a similar fashion as the sole softest \pyrCA in row \twotwoone of \cref{fig: hcp}.
Moreover, a gradual transition from three separate families into a combined higher-multiplicity family results when the slip resistance ratios approach one, as observed in the progression from row \onetwofour toward \onetwotwo and \oneoneone of \cref{fig: hcp}, which strongly diminishes the surface effect.
Conversely, an increasing slip resistance ratio between the easiest and harder slip families decreases the likelihood of activation for the harder families, resulting in a widened span of slip family activity, particularly at the surface (compare row \oneonetwo and \oneonefour in \cref{fig: hcp}).

Comparing the behavior between HCP and BCC (at slip resistance ratios of 2), the least and most active slip families differ by more than four orders of magnitude for BCC but no more than two orders of magnitude for HCP (\cf rows \onetwo and \twoone in \cref{fig: bcc} and rows \onetwotwo, \twotwoone, and \oneonetwo in \cref{fig: hcp}).
This discrepancy is rooted in the similarity of the yield surfaces associated with the two BCC slip families in contrast to the dissimilarity of those associated with the \basalA, \prismA, and \pyrCA slip families \citep{10.1016/0001-6160(85)90025-2}.
Since the smallest rotation to align a \hkl{110} slip system with the nearest \hkl{112} system is only \ang{30}, the two yield surfaces are very similar and, therefore, even a small difference in slip resistances between those two families strongly favors the easier family under virtually all possible deformation conditions.
In contrast, a large misorientation of \ang{64} and \ang{79} between a \pyrCA slip system and its nearest \basalA and \prismA slip system, respectively, implies that \basalA---and even more so \prismA---offers kinematic degrees of freedom where \pyrCA is lacking, despite its high multiplicity of 12.
This kinematic anisotropy of \pyrCA causes the \pyrCA yield surface to be relatively extended (harder to reach) in those directions where \basalA and \prismA are most facile.
Consequently, a very large slip resistance of \basalA and \prismA relative to \pyrCA would be required to fully preclude their activity under deformation conditions unfavorable for \pyrCA, \ie, where the \pyrCA yield surface is extended.
This requirement of relatively large slip resistance ratios explains why, despite \pyrCA being \emph{the} easiest family in row \twotwoone and \emph{one of} the easiest in row \oneoneone of \cref{fig: hcp}, there is still appreciable activity of \basalA and especially \prismA, since both contribute slip activity for deformation conditions that do not align well with \pyrCA kinematic degrees of freedom.

Another consequence of the similarity between the two BCC slip families, in contrast to the dissimilar HCP families, is the lack of a large (relative) surface effect for the hardest family in BCC compared to a much more substantial effect in HCP (\pyrCA in row \oneonefour of \cref{fig: hcp}).
This is due to the shape of the yield surfaces spanned by each of the two BCC slip families being very similar, so any changes in the deformation boundary conditions experienced near the surface compared to the bulk will not substantially alter the ensuing slip activity, as demonstrated by the modest surface effects observable in \cref{fig: bcc} (particularly row \oneone).

\section{Conclusions}

Published slip resistance values of, for instance, magnesium and commercially pure titanium, determined through a variety of experimental means exhibit differences that suggest a systematic influence of the surface on the activity of hexagonal slip families.
Crystal plasticity simulations of polycrystalline film and bulk structures were carried out to investigate the influence of a surface on the slip family activity in hexagonal close-packed (HCP) and body-centered cubic (BCC) materials.
We could demonstrate that two main factors determine the strength of a surface effect, namely the contrast in slip family resistances (slip resistance ratios) and how similar the different slip families are (with similarity generally increasing with increasing multiplicity).
Any variation in elastic anisotropy had no appreciable influence on the surface effect.
Compared to the bulk interior, the activity of harder slip families always diminishes near a surface.
The activity of the easiest slip family, especially when it has less than about six available slip systems, exhibits a notable increase in activity near the surface due to the relaxed constraints.
Such a surface effect extends to a depth of approximately one to two average grain diameters, and it is amplified with increasing slip resistance ratios.

Based on the present results, given that \pyrCA slip is generally more difficult than \basalA or \prismA slip in HCP materials, a significant surface effect  of \pyrCA should be anticipated.
Specifically, if the sample dimension or slip resistance measurement methodology restricts the probed depth to less than about two average grain diameters, a significant overestimation of the slip resistance for the hard \pyrCA family is expected, along with an underestimation for the easiest family.
The magnitude of this surface effect increases with the slip resistance ratio between the hardest and softest families, and inversely with the multiplicity of the easiest family.
For HCP metals, the shallower the probed depth, and the greater the ratio between hardest (\pyrCA) and easiest (\basalA or \prismA) slip resistance, the stronger the expected surface effect.

In contrast, combinations of high-multiplicity slip families do not exhibit a significant surface effect, as demonstrated using \hkl{110} and \hkl{112} slip families in BCC materials.
The observed larger average activity of \hkl{112} over \hkl{110} slip for equal resistance and multiplicity is connected to their kinematic difference (twice as many \hkl{112} than \hkl{110} planes) that makes it slightly more likely to have the highest resolved shear stress on a \hkl{112} slip system.

To sum up, this investigation reveals that interpretation of \emph{statistical} slip behavior needs to be done carefully, as responses collected near a surface could differ from bulk locations.
This is particularly relevant for materials featuring low-multiplicity slip families such as HCP, whereas BCC and FCC metals, in which all slip families have high multiplicity, are virtually unaffected.

\section*{Data availability}

Scripts employed to generate the data that support the findings of this study are available at \url{https://github.com/CathyBing/slip_surface_effect}.

\section*{Acknowledgment}

This study was supported by the U.S. Department of Energy, Office of Science, Office of Basic Energy Sciences, under award numbers DE-SC0001525 and DE-SC0009960 as well as by Michigan State University through computational resources provided by the Institute for Cyber-Enabled Research.
We appreciate the input from our colleague Dr.~Eric M. Taleff, University of Texas at Austin, and insightful comments by one of the reviewers that led to significant improvements in the scope and clarity of presentation of this work.

\bibliographystyle{elsarticle-num-names}
\bibliography{Bing_etal2023}

\end{document}